\documentclass[conference]{IEEEtran}
\IEEEoverridecommandlockouts
\usepackage{cite}
\usepackage{float}
\usepackage{amsmath,amssymb,amsfonts}
\usepackage{algorithmic}
\usepackage{graphicx}
\usepackage{textcomp}
\usepackage{xcolor}
\def\BibTeX{{\rm B\kern-.05em{\sc i\kern-.025em b}\kern-.08em
    T\kern-.1667em\lower.7ex\hbox{E}\kern-.125emX}}

\usepackage{xurl}

\usepackage{microtype}
\usepackage{booktabs} % for professional tables
\usepackage{hyperref}

\usepackage{amsmath}
\usepackage{amssymb}
\usepackage{mathtools}
\usepackage{amsthm}
\usepackage{cite}
\usepackage[caption=false,font=footnotesize]{subfig}

\usepackage[capitalize,noabbrev]{cleveref}

\theoremstyle{plain}

\theoremstyle{definition}

\theoremstyle{remark}

\makeatletter
\newcommand*\bigcdot{\mathpalette\bigcdot@{.5}}
\newcommand*\bigcdot@[2]{\mathbin{\vcenter{\hbox{\scalebox{#2}{$\m@th#1\bullet$}}}}}
\makeatother

\usepackage[textsize=tiny]{todonotes}
\usepackage{enumitem}
\usepackage{soul}
\usepackage{graphicx}    % For images
\begin{document}

\title{Position Paper: Emergent Machina Sapiens Urge Rethinking Multi-Agent Paradigms in Critical Infrastructures
\\
% {\footnotesize \textsuperscript{*}Note: Sub-titles are not captured in Xplore and should not be used}
\thanks{This work is supported in part by the Concordia University Research Chair program, the Natural Sciences \& Engineering Research Council (NSERC) of Canada under grants RGPIN-2018-06724 and RGPIN-2025-05097, and the U.S. National Science Foundation under the grant number 2330504. %, the ... and the ... . %The authors would also like to thank ..., ..., and ... for their valuable feedback. 
}
}

% \twocolumn[
% \icmltitle{Position: Emergent Machina Sapiens Urge Rethinking Multi-Agent Paradigms}




% \begin{icmlauthorlist}
% \icmlauthor{Hepeng Li}{equal,yyy}
% \icmlauthor{Yuhong Liu}{equal,comp}
% \icmlauthor{Jun Yan}{sch}
% % \icmlauthor{Firstname4 Lastname4}{sch}
% % \icmlauthor{Firstname5 Lastname5}{yyy}
% % \icmlauthor{Firstname6 Lastname6}{sch,yyy,comp}
% % \icmlauthor{Firstname7 Lastname7}{comp}
% % %\icmlauthor{}{sch}
% % \icmlauthor{Firstname8 Lastname8}{sch}
% % \icmlauthor{Firstname8 Lastname8}{yyy,comp}
% \end{icmlauthorlist}

% \author{Anonymous Authors}
\author{
\IEEEauthorblockN{Hepeng Li\IEEEauthorrefmark{1},
Yuhong Liu\IEEEauthorrefmark{2},
Jun Yan\IEEEauthorrefmark{3},
Jie Gao\IEEEauthorrefmark{4},
Xiao'ou Yang\IEEEauthorrefmark{2}}
\IEEEauthorrefmark{1}Department of Electrical \& Computer Engineering, University of Maine, Orono, ME, USA\\
\IEEEauthorrefmark{2}Department of Computer Science \& Engineering, Santa Clara University, Santa Clara, CA, USA\\
\IEEEauthorrefmark{3}Concordia Institute for Information Systems Engineering, Concordia University, Montr\'eal, QC, Canada\\
\IEEEauthorrefmark{4}Department of Transport \& Planning, Delft University of Technology, Delft, Netherlands\\
Email: hepeng.li@maine.edu;
yhliu@scu.edu;
jun.yan@concordia.ca;
j.gao-1@tudelft.nl;
xyang13@scu.edu
}

% \author{
% \IEEEauthorblockN{Hepeng Li}
% \IEEEauthorblockA{
% % \textit{Department of Electrical \& Computer Engineering} \\
% \textit{University of Maine}\\
% Orono, ME, United States\\
% hepeng.li@maine.edu}
% \and
% \IEEEauthorblockN{Yuhong Liu}
% \IEEEauthorblockA{
% % \textit{Department of Computer Science \& Engineering} \\
% \textit{Santa Clara University}\\
% Santa Clara, CA, United States\\
% yhliu@scu.edu}
% \and
% \IEEEauthorblockN{Jun Yan}
% \IEEEauthorblockA{
% % \textit{Concordia Institute for} \\
% % \textit{Information Systems Engineering} \\
% \textit{Concordia University}\\
% Montr\'eal, QC, Canada\\
% jun.yan@concordia.ca}
% \and
% \IEEEauthorblockN{Jie Gao}
% \IEEEauthorblockA{
% % \textit{Department of Transport \& Planning} \\
% \textit{Delft University of Technology}\\
% Delft, Netherlands\\
% j.gao-1@tudelft.nl}
% \and
% \IEEEauthorblockN{Xiao'ou Yang}
% \IEEEauthorblockA{
% % \textit{Department of Mechanical Engineering} \\
% \textit{Santa Clara University}\\
% Santa Clara. CA, United States\\
% xyang13@scu.edu}
% }
\maketitle

% \icmlaffiliation{yyy}{Department of Electrical \& Computer Engineering, University of Maine, Orono, USA}
% \icmlaffiliation{comp}{Santa Clara University, Computer Science and Engineering, Santa Clara, USA}
% \icmlaffiliation{sch}{Concordia Institute for Information Systems Engineering, Concordia University, Montreal, Canada}

% \icmlcorrespondingauthor{Jun Yan}{jun.yan@concordia.ca}


% \vskip 0.3in
% ]




\begin{abstract}
Artificial Intelligence (AI) agents capable of autonomous learning and independent decision-making hold great promise for addressing complex challenges across various critical infrastructure domains, including transportation, energy systems, and manufacturing. However, the surge in the design and deployment of AI systems, driven by various stakeholders with distinct and unaligned objectives, introduces a crucial challenge: How can uncoordinated AI systems coexist and evolve harmoniously in shared environments without creating chaos or compromising safety? To address this, we advocate for a fundamental rethinking of existing multi-agent frameworks, such as multi-agent systems and game theory, which are largely limited to predefined rules and static objective structures. We posit that AI agents should be empowered to adjust their objectives dynamically, make compromises, form coalitions, and safely compete or cooperate through evolving relationships and social feedback. Through two case studies in critical infrastructure applications, we call for a shift toward the emergent, self-organizing, and context-aware nature of these multi-agentic AI systems.

% The predesigned objectives, which are to be learned by these rationally designed and self-serving agents, are unknown to other agents who are also learning to optimize their own strategies. 

% Uncoordinated design

% Uncoordinated learning

% Calls for a higher inter-agentic framework

% Assess their decision and learning quality in the ecosystem
% Coordinate the Update learning objectives of serve-serving, continuous learning, and highly rational agents toward unknown common grounds.

% Reveal and mitigate potential conflicts, collisions, and other chaotic behaviors by spontaneously identifying and updating beyond the initial design objectives.

% Enable solutions that autonomously update the agents' rationality models, mutual relationships, and objective functions that are drifting over time after deployment.

% In return, guide the development of individual but interoperable AI solutions in the fast-expanding multi-stakeholder ecosystem.

% This can result in conflicting, chaotic, and/or colluded interactions learned by these autonomous agents post-deployment, which are unpredicted or undesirable by the AI developers in the conventional centralized design paradigm.

% new effort that address interoperability 

% move beyond

% address from different angle

% learn to coordinate from themselves
\end{abstract}


\section{Introduction}
\label{intro}
% definition of Agentic AI
%Recently, the rapid advancement of generative artificial intelligence (AI) technologies has paved the way for agentic AI, which refers to AI systems designed to pursue goals in a highly autonomous manner that resembles human-like agencies \cite{}. Because of AI agents' high autonomy, they can greatly reduce the required human involvement in achieving the desired goals. However, it also indicates that setting the initial goals becomes the only time/place where a human user can intervene. Such limited interactions start to raise increasing trust concerns on whether AI agents' decisions and actions can always align with the designer/owner's values, especially when most AI agents lack transparency, explainability, and traceability (e.g., effective measurements). More importantly, when more heterogeneous agentic AIs representing different stakeholders with various goals are deployed in a shared environment, will their goals conflict with one another? Can they adaptively learn not only from a more dynamic environment but also from other highly adaptive peers while still achieving their initial goals? Will an AI agent establish relationships with other agents? Will such relationships evolve and even influence an agent's future behaviors? How can human users observe/measure, trust, or even intervene to avoid high risks in these complex scenarios?   
% Enormous and rapidly growing application scenarios of agentic AI 

The advancement in autonomous AI systems, such as deep reinforcement learning \cite{mnih2015human,SilverHuangEtAl16nature}, agentic artificial intelligence \cite{shavit2023practices,10.1145/3593013.3594033}, self-supervised learning \cite{SSL,chen2020big}, meta-learning \cite{pmlr-v70-finn17a,NEURIPS2019_072b030b}, etc., has ushered in a new era of machina sapiens, where machines transcend their traditional roles to become independent learners, problem-solvers, and human assistants and collaborators. The emergence of these autonomous systems promises transformative capabilities in addressing complex tasks across various critical infrastructure domains \cite{luitel2014cellular}, such as smart transportation \cite{ault2021reinforcement}, smart grid and energy systems \cite{8521585,9721402}, and robotics and smart manufacturing \cite{pmlr-v87-kalashnikov18a,tang2024deepreinforcementlearningrobotics,zhou2020deep,zhang2022dynamic}, among others.

With the enormous potential of AI adoptions, we envision an ecosystem where different types of autonomous systems, designed and deployed independently by various stakeholders with distinct goals, will interact, evolve, and coexist as \textit{agents} \cite{10.5555/3312046}. Autonomous cars from multiple manufacturers will share the same roads, while intelligent energy management systems deployed by utilities and prosumers will co-manage the power grid. Additionally, robotic assistants customized for diverse user needs will collaborate in shared spaces, such as manufacturing facilities, hospitals, warehouses, and homes. However, in such environments, agents may find it challenging to achieve their goals alongside other AI agents if they adhere solely to the rules and objectives as designed. The underlying conflicts, competition, and misalignment in the goals and behaviors could lead to failures at the societal level, which in turn undermine the success of individual agents, causing chaotic disruptions that compromise the safety, reliability, and ethical integrity of real-world AI applications.

The interconnected agentic ecosystems raise critical questions: How can AI agents with different goals and architectures coevolve and collaborate in shared environments? How can they adapt to unforeseen situations without jeopardizing system-wide safety and performance? These challenges extend beyond technical coordination. Autonomous systems, despite being designed and deployed independently, must nonetheless learn to cooperate, negotiate trade-offs, and adhere to societal rules that balance individual objectives with collective well-being. For instance, should a delivery robot prioritize speed over safety when sharing sidewalks with humans? How should an autonomous car handle conflicting interests between its passengers and other road users? As the diversity and ubiquity of AI systems grow, these questions become increasingly urgent.

Traditional frameworks such as multi-agent reinforcement learning \cite{Ming1997,Claus1998,Kar2013,lowe2020multiagent,zhang2019multiagent} and game theoretical methods \cite{littman1994markov,pmlr-v202-slumbers23a,mao2024alympics,yang2024exploring} have been widely used to model interactions among agents, but they often fall short when applied to such open environments \cite{gal2022multi} for several reasons:

\textit{$\bigcdot$\ Static Objectives and Equilibria:} Existing approaches often assume AI agents optimize with fixed utility or reward structures toward static equilibria. These assumptions do not hold in systems where contexts shift unpredictably and common interests change over time, requiring agents to be capable of dynamically adjusting their goals post deployment.

\textit{$\bigcdot$\ Predefined Rules for Interactions:} Conventional frameworks typically model AI interactions based on predefined relationships. However, these interactions may need to dynamically shift between cooperation and competition, depending on the evolving contexts and objectives.

\textit{$\bigcdot$\ Pre-Designed Coordination Mechanisms:} Most learning frameworks rely on the meticulous design of coordination algorithms in advance to ensure consensus or convergence, which is impractical in an ecosystem where AI agents are designed and deployed independently.

\textit{$\bigcdot$\ Curse of Scalability:} Coordinating and aligning the behaviors of numerous AI agents demands resource-intensive setups, leading to exponential increases in the complexity of computation and interaction. This poses significant challenges for traditional paradigms in large-scale applications.

These limitations are fundamentally rooted in paradigms that prioritize the design of multi-agent interactions for coordination rather than permitting interactions to emerge organically among AI agents. The advent of machine intelligence presents a transformative opportunity to transcend these constraints with innovative approaches to managing the dynamic, emergent, and socially embedded complexities in the emergent AI ecosystems.

\textbf{Our Position:} We posit that the growing trend of agentic AI development demands a fundamental rethinking of how we conceptualize % and stimulate autonomy and 
interoperability among an influx of AI agents. Instead of relying on pre-engineered rules and static reward structures, agents should possess the flexibility to adapt to the diverse and evolving world. This flexibility involves the capacity to align their individual goals with broader systems to balance their objectives with collective considerations, such as safety, fairness, and ethical standards. By fostering such autonomy, agents can not only coexist but also co-thrive in open-ended environments, which would otherwise challenge existing paradigms.

Realizing this vision entails empowering agents to adapt their learning mechanisms, negotiate trade-offs, form coalitions, and engage safely in cooperation or competition. AI agents must be capable of proactively collecting feedback and critics by building trust in dynamic relationships and interactions with peers. This paradigm prioritizes adaptability, harmony, and resilience, shifting from pre-engineered frameworks to self-organizing and context-aware systems. To this end, we advocate for the development of novel methodologies and design principles for multi-agentic ecosystems that foster cooperation among AIs and between AI and humans, ensuring that the benefits of AI align with societal values.

%\textbf{Contributions:} 
Our position aims to contribute by: %
%In this paper, we make the following contributions to multi-agent systems and agentic AI:
\vspace{-0.3em}
\begin{itemize}[leftmargin=.11in]
\item \textbf{Inspiring Future Research Directions:} We highlight the opportunities and challenges posed by the increasing prevalence of autonomous, diverse, and independently developed AI agents in shared environments. We encourage the research community to explore innovative methodologies by reexamining existing paradigms while addressing the complexities of emergent AI ecosystems.
\item \textbf{Proposing a Framework for Dynamic Interaction:} We advocate for a new framework that integrates dynamic norms and adaptive protocols as fundamental components for governing multi-agent interactions. This framework equips agents to continuously evolve their behaviors based on feedback, fostering alignment, coordination, and resilience in diverse, open-ended environments.
\item \textbf{Grounding Challenges in Real-World Contexts:} We situate the theoretical challenges of emergent agentic ecosystems within practical scenarios. This contextualization underscores the necessity of adaptive and dynamic approaches while demonstrating how the proposed framework can accommodate real-world complexities.
\end{itemize}
% \medskip

\section{The Ecology of An Agentic AI Ecosystem}

\subsection{The Nature and Nurture of Autonomous Agents}
Analyzing how autonomous agents develop and perform in multi-agent settings is helpful in distinguishing between nature and nurture. Although these terms may evoke biological analogies, we use them here to articulate two complementary facets of agent behavior: nature is relatively fixed at inception, and nurture evolves as circumstances change.

From a \textbf{nature} perspective, each agent begins with an intrinsic capacity to:

\begin{itemize}%\setlength\itemsep{0.02em}
    \item Evolve greedily in a goal-driven manner
    \item Learn autonomously from interactive feedback
    \item Optimize quantitatively with precise metrics
\end{itemize}

These built-in qualities establish a stable foundation: agents know what they are striving for, have independent means of adapting their strategies, and can precisely measure outcomes relevant to their predefined goals.

Yet, as no environment remains entirely static, agents must also embody a \textbf{nurture} component, where

\begin{itemize}%\setlength\itemsep{0.02em}
    \item Goal may shift by contextual and dynamic factors
    \item Learning can be influenced by peer agents or humans
    \item Metrics may be uncertain or incompletely defined
\end{itemize}
Under these nurture influences, the pursuit of a fixed objective becomes a nuanced negotiation of evolving factors.

This duality sheds light on how agents develop over time. They begin with certain high-level design choices, learning algorithms, and objective functions that guide their initial behavior. However, when confronted with real-world change, they adapt: goals may be reassessed; unanticipated signals can refocus their decisions; and newly emerging metrics can upend earlier assumptions about optimal performance. In short, nature ensures that agents have consistent, goal-driven architectures, while nurture ensures they do not stagnate but continually revise their internal parameters in light of external developments.

When many such agents, each guided by distinct but similarly structured nature-and-nurture dynamics, coexist in a shared environment, interactions can become unexpectedly intricate. Only by first recognizing how an agent's built-in qualities (nature) intersect with its adaptive growth (nurture) can we fully grasp why multi-agent ecosystems exhibit both boundless potential and inherent challenges.

\subsection{Alternative Positions and Opposing Views}
% While this paper advocates for a rethinking of multi-agent systems through the lens of adaptability, collaboration, and emergent alignment, 
Several existing approaches challenge or diverge from our position, which offer meaningful contributions but may fall short of addressing the full complexity and dynamism of real-world, uncoordinated AI ecosystems.

\paragraph{Agentic AI: A Focus on Individual Autonomy.}
Agentic AI \cite{wooldridge1995intelligent,durante2024agentaisurveyinghorizons,shavit2023practices}, including agentic AI workflows \cite{zhang2024aflow}, emphasizes the development of highly autonomous agents capable of independently learning to achieve their objectives with minimal external guidance. This approach prioritizes self-sufficiency, advanced decision-making capabilities, and optimization of individual goals, often relying on techniques like reinforcement learning, meta-learning \cite{yun2023quantum}, or hierarchical policies. Sufficiently advanced agentic AI could overcome many coordination challenges by developing internal mechanisms to handle conflicts, adapt to changes, and negotiate with peers.

While agentic AI excels at individual autonomy, its emphasis on independent optimization introduces challenges in multi-agent settings. When each agent optimizes for its own objectives without shared guiding principles, misalignment can emerge at the system level, particularly in environments where collective welfare is critical. Agents may compete for resources, disrupt one another's plans, or fail to recognize mutually beneficial opportunities due to their isolated learning processes. Relying solely on individual agents to resolve coordination challenges risks fragmented and unpredictable behaviors, which may be difficult to regulate or steer toward desirable outcomes in open-ended environments.

\paragraph{Context-Aware Multi-Agent Systems.}
Context-aware multi-agent systems have long been explored as a means of enhancing coordination and adaptability in multi-agent environments \cite{4197340,du2024surveycontextawaremultiagentsystems}. These approaches use context modeling to enhance agents' ability to perceive, interpret, reason, learn, and adapt to environmental and situational factors for optimized decision-making. By integrating information such as task dependencies, resource availability, and agent roles, they enable structured adaptation, aligning agent behaviors with system-wide objectives while maintaining flexibility in uncertain settings. Many approaches incorporate explicit rules \cite{jelen2022multi}, consensus mechanisms \cite{amirkhani2022consensus}, shared societal expectations \cite{mckee2021multi}, or structured/learnable communication protocols \cite{sukhbaatar2016learning,10.5555/3295222.3295385} to help agents make informed decisions while minimizing conflicts.% This structured adaptation has been particularly successful in domains where agents benefit from having a common understanding of environmental constraints and shared goals.

Context-aware multi-agent systems often rely on engineered coordination mechanisms, where context parameters are pre-encoded. However, emerging AI societies involve agents that may have divergent goals and must learn to interact without relying on pre-engineered contextual models. In such cases, rigid context-aware mechanisms may limit agents' ability to dynamically shape their own interactions. Moreover, these systems typically require centralized or semi-centralized coordination, such as explicit goal alignment or externally provided context. While these mechanisms enhance predictability, they may restrict the emergence of organic relationships and self-organized norms that allow agents to develop truly adaptive strategies in open environments. This limitation becomes particularly evident when agents must not only adapt to a static context but also shape and redefine the context itself through their interactions.

\section{A Multi-Agentic AI Framework}
\subsection{Components of the Framework}
The proposed framework is built around the idea of a dynamic ecosystem of AI agents. These agents act, react, and evolve in a world of diverse environments to pursue their objectives while being subject to environmental and societal boundaries. The framework comprises three key components: \textbf{agents}, \textbf{world}, and \textbf{norms}.
\begin{figure}[ht]
% \vskip 0.2in
\begin{center}
\centerline{\includegraphics[width=0.9\columnwidth]{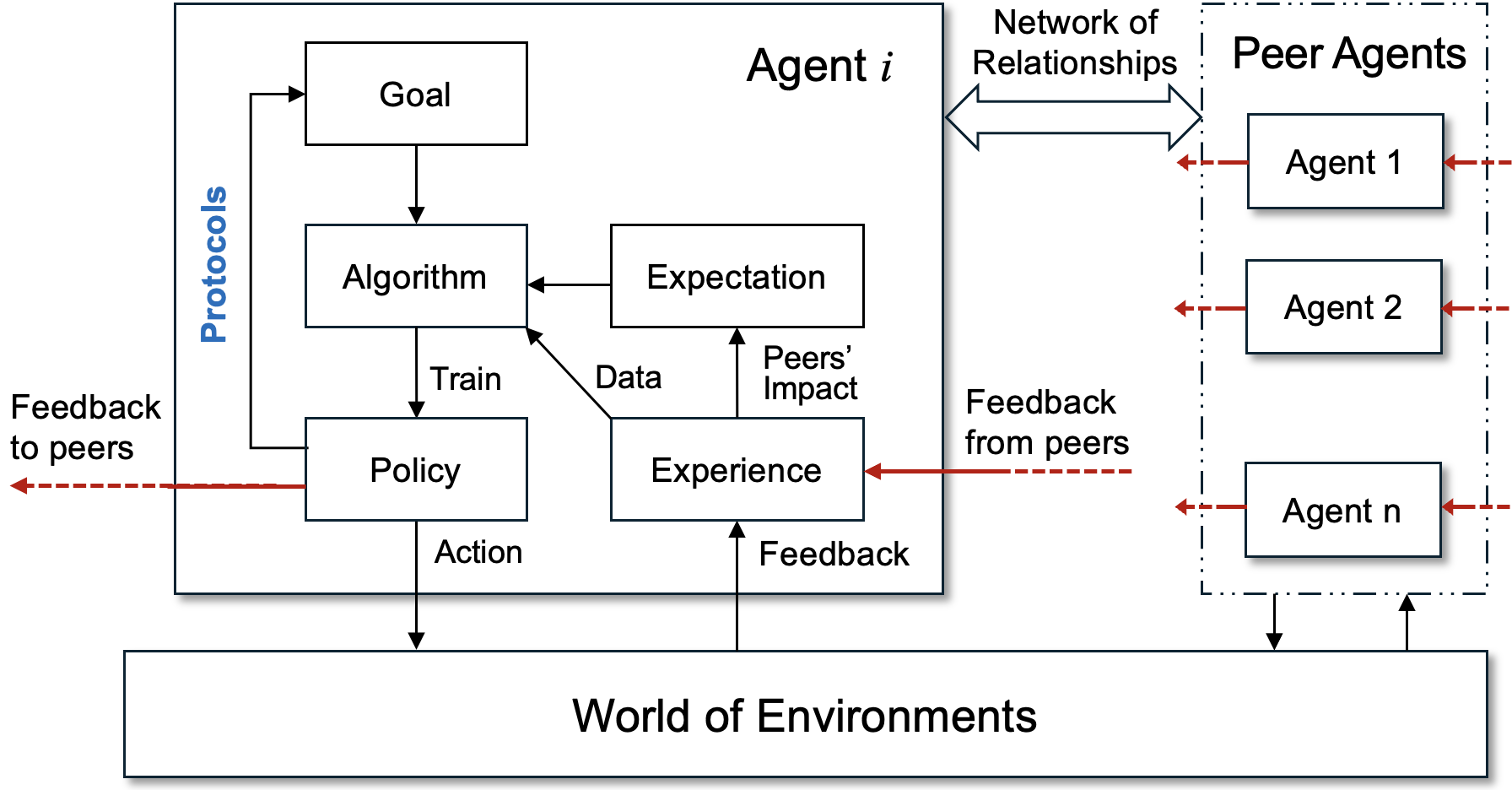}}
\caption{The framework enables agents to refine their goals, policies, and expectations through feedback from both the environment and peer agents. Protocols shape interactions by dynamically adjusting goals, evolving based on societal impacts, peer expectations, and relationship networks.}
\label{framework}
\end{center}
\vskip -0.3in
\end{figure}

\noindent\textbf{Agents}: Agents are social, intelligent entities that act autonomously, learn from experience, and interact with others. An agent is defined as $A:=\langle J, \pi, \mathcal{O}, \mathcal{B}, \mathcal{G}\rangle$, where 
$J$ defines the \textit{Goal}, i.e., the agent's objective or purpose. 
$\pi$ is the \textit{Policy}, the rule for acting and communicating. $\mathcal{L}$ is \textit{Algorithm}, the method the agent uses to learn. 
$\mathcal{B}$ is \textit{Experience}, the agent's accumulated knowledge.
$\mathcal{G}$ is \textit{Relationships}, the agent's social connections.
    %\item $\delta$ -- \textbf{Pace}: The act or communication intervals.
% \begin{itemize}
%     \item $J$ -- \textbf{Goal}: The agent's objective or purpose.
%     \item $\pi$ -- \textbf{Policy}: The rule for acting and communicating.
%     \item $\mathcal{L}$ -- \textbf{Algorithm}: The method the agent uses to learn.
%     % \item $\mathcal{O}$ -- \textbf{Perception}: The method the agent uses to learn.
%     \item $\mathcal{B}$ -- \textbf{Experience}: The agent's accumulated knowledge.
%     \item $\mathcal{G}$ -- \textbf{Relationships}: The agent's social connections.
%     %\item $\delta$ -- \textbf{Pace}: The act or communication intervals.
% \end{itemize}
%This definition is intentionally abstract to capture the most fundamental features shared by autonomous AI and humans. However, agents can differ in each component, enabling the creation of a highly diverse and heterogeneous multi-agent system. For instance, agents may operate at different paces depending on their tasks and goals, necessitating synchronization for effective cooperation. Another unique feature is the inclusion of dynamic relationships $\mathcal{G}$. While agents operate independently, their relationships evolve over time, allowing them to interact based on social norms.

\noindent\textbf{World}: The world consists of a variety of environments $\mathcal{W}=\{E_{1},\dots,E_{m}\}$, with which agents can interact individually or simultaneously. Each environment in the world is defined as a tuple $E:=\langle S, A, P, R\rangle$, where
$S$ defines the \textit{States}, i.e., a set of the environment's possible states, which may not be fully observable by the agents.
$A$ is \textit{Actions}, a set of external inputs that influence the environment dynamics.
$P$ is \textit{Dynamics}, the rules governing how the environment evolves in response to actions or external factors.
$R$ is \textit{Reward}, the immediate feedback provided by the environment based on agents' actions.
    % \item $\ t$ -- \textbf{Timescale}: The temporal framework, which could be discrete, continuous, or event-triggered.

% \begin{itemize}
%     \item $S$ -- \textbf{States}: A set of the environment's possible states, which may not be fully observable by the agents.
%     \item $A$ -- \textbf{Actions}: A set of external inputs that influence the environment dynamics.
%     \item $P$ -- \textbf{Dynamics}: The rules governing how the environment evolves in response to actions or external factors.
%     \item $R$ -- \textbf{Reward}: The immediate feedback provided by the environment based on agents' actions.
%     % \item $\ t$ -- \textbf{Timescale}: The temporal framework, which could be discrete, continuous, or event-triggered.
% \end{itemize}
%Environments may also be interconnected, with dependencies or hierarchical relationships. For example, the dynamics of a local environment can cascade to influence a larger, overarching system, enabling complex, multi-layered interactions.

\noindent\textbf{Norms}: Norms represent the societal structures and guiding principles that govern agent interactions and relationships. It is defined as a tuple $M:=\langle \mathcal{N}, \mathcal{I}, \mathcal{E}, \mathcal{P}\rangle$, where
$\mathcal{N}$ defines the \textit{Networks}, i.e., the structure of agents' relationships.
$\mathcal{I}$ is \textit{Impacts}, the influence of agents in a network on the environments and other agents.
$\mathcal{E}$ is \textit{Expectations}, the beliefs agents hold about the likely behaviors or outcomes of others.
$\mathcal{P}$ is \textit{Protocols}, the rules or standards that regulate how agents establish relationships and interact.

\subsection{Connecting Norms, Agents, and the World}

\noindent\textbf{Evolution of Norms.}
Norms are the backbone of the proposed framework, shaping the interactions, behaviors, and learning mechanisms of agents.  Norms can be classified into two types: \textbf{fixed norms}, which establish a static structure with predetermined rules, e.g., MARL and games, and \textbf{dynamic norms}, which evolve through agent interactions and feedback to enable adaptive and self-organized relationship building. We primarily focus on dynamic norms.

The evolution of norms begins with initial protocols, $\mathcal{P}=\{\mathcal{P}_{1},\dots,\mathcal{P}_{n}\}$, that establish rules for agents to build cooperation or manage conflicts. For instance, a basic protocol, $\mathcal{P}_{i}=\alpha \cdot H_i$, may dictate resource-sharing mechanisms by discouraging hoarding as follows:
\begin{equation}
\label{protocol_i}
    J_i^{\mathcal{P}} = J_i - \mathcal{P}_{i}, \forall i
    \vspace{-0.2em}
\end{equation}
where $J_i$ is agent $i$'s original goal (e.g., selfishly hoarding), $J_i^{\mathcal{P}}$ is the goal subject to the protocol $\mathcal{P}_{i}$, and $\alpha$ is the penalty coefficient for the hoarding behavior metric $H_i$.

Protocols can emerge as penalties on agents' objectives, as shown in (\ref{protocol_i}), incentivizing agents to adjust their goals to align with collective priorities. %This approach effectively models various forms of interaction, such as cooperation, competition, and hierarchical structures. This adjustment also enables agents to adapt to different tasks dynamically. 
Alternatively, protocols may act as constraints that limit the actions of agents, ensuring that their individual objectives do not violate safety, fairness, or ethical boundaries. This is particularly useful in applications like autonomous vehicles.

%Through the proposed framework, it is now possible for protocols to be learnable and evolve. The evolution of protocols can only be enabled if the agents' societal impacts $\mathcal{I}$ and the expectations $\mathcal{E}$ held by peers could be reflected on the changes of the agents' relationship network $\mathcal{N}$, which is then fed back to the agents for them to evaluate how it will help/prevent them from achieving their goals. Only when such a feedback loop is activated can agents learn whether their protocols should be reinforced or replaced. For example, an agent's initial protocol may be cooperating with other agents. Following this initial protocol, suppose this agent consistently meets or exceeds the expectations of its peers; it will observe whether such protocol would lead to enhanced trust or higher influence and, consequently, a higher probability of achieving its goals. If yes, such protocol will be reinforced. Otherwise, the protocol may be refined or replaced. On the other hand, assume an agent's initial protocol is to be selfish and ignore all other agents' expectations. Following this protocol, this agent may consistently take advantage of other agents. Assume in an environment where collaboration among agents is critical for their success; this agent may end up with significantly fewer opportunities to participate in collaborations, making it very unlikely to achieve its own goals. Such feedback may lead to the protocol to be revised or replaced.

Protocols are learnable and evolve over time, helping agents achieve individual goals by managing conflicts, developing cooperation, or delegating subtasks to other agents. The evolution of protocols could be influenced by the agents' relationship network $\mathcal{N}$, their societal impacts $\mathcal{I}$, and the expectations $\mathcal{E}$ held by peers. When agents consistently meet or exceed the expectations of their peers, they reinforce existing protocols, fostering trust and enhancing their influence. Conversely, deviations from expected behaviors—whether due to conflicts, inefficiencies, or environmental shifts—can trigger the refinement or replacement of protocols.

The relationship network $\mathcal{N}$, the impacts $\mathcal{I}$, and the expectations $\mathcal{E}$ are not static. They evolve as agents interact with each other. As agents learn more about their environment and the behaviors of others, their actions and the resulting impacts shift, changing how they are perceived and how they perceive others. These dynamic relationships guide agents toward more effective adaptation. Through this continuous evolution of $\mathcal{N}$, $\mathcal{I}$, and $\mathcal{E}$, agents gradually refine their strategies, norms, and interactions, aiming to achieve their goals in a dynamic ecosystem.

The stability of protocols can be analyzed using a dynamical system representation:
\begin{equation}
    \frac{d\mathcal{P}}{dt} = -\gamma \cdot (\mathcal{P} - \mathcal{P}^*),
    \vspace{-0.2em}
\end{equation}
where \(\gamma\) is a stability coefficient and \(\mathcal{P}^*\) is the equilibrium protocol. A special case is that the equilibrium optimizes collective outcomes: $\mathcal{P}^* = \arg \max_{\mathcal{P}} \sum_{i} J_i(\pi_i, \mathcal{P})$,
where $\mathcal{P}^*$ can be seen as protocols used for traditional cooperative MARL problems. Note that protocols may not necessarily converge, indicating an ever-evolving ecosystem.

\paragraph{Evolution of Relationships.}
The relationships among all agents collectively form the network $\mathcal{N}$, serving as the social fabric of the framework. Agent $i$'s relationships can be represented by a directed weighted graph $\mathcal{G}_{i}$, where each edge $g_{ij} \in \mathcal{G}_{i}$ originates from agent $i$ and points to agent $j$, with a weight $w_{ij}$, indicating the strength and nature of a relationship. The relationships are not necessarily symmetric.

The relationship strength $w_{ij}$ is dynamic and evolves over time based on interactions and feedback, updated by:
% \vspace{-0.5em}
\begin{equation}
w_{ij} \leftarrow f(w_{ij}, \mathcal{I}_{ij}, \mathcal{E}_{ij})
\vspace{-0.2em}
\end{equation}
where $f(\cdot)$ is a function that governs the update process, $\mathcal{I}_{ij}$ represents the impact of agent $j$ observed by agent $i$, and $\mathcal{E}_{ij}$ denotes the expectation that agent $i$ holds about agent $j$. %The number of edges in the graph $\mathcal{G}_{i}$ changes dynamically as agent 
$i$ establishes or dissolves relationships.

The evolution of relationships creates a dynamic network, allowing agents to form coalitions, dissolve unproductive ties, and adapt their connections to establish new norms. These adjustments create a feedback loop where successful collaborations strengthen relationships, while conflicts prompt reevaluation and adaptation. For example, agents with aligned goals may form coalitions to pool resources or address complex tasks. Mathematically, a coalition $C$ may form when: $\sum_{i, j \in C} w_{ij} \geq \theta$,
where $\theta$ is a threshold for the minimum collective relationship strength required for coalition stability. Additionally, agents may prioritize communications or knowledge sharing with trusted peers, reducing resource expenditure and enhancing overall efficiency. This prioritization could be modeled as: $P_i(j) = \frac{w_{ij}}{\sum_k w_{ik}}$, where $P_i(j)$ is the probability that agent $i$ communicate with $j$.

\paragraph{Cross-Environment Interactions.} In our framework, agents interact across multiple environments, e.g., a hierarchy or heterarchy of energy management systems, each governed by unique rules, dynamics, and challenges. Unlike traditional frameworks confined to a single shared environment, these environments may be interconnected, overlapping, or nested within hierarchical structures, allowing agents to influence and be influenced by actions across environments.

Consider agents $i$ and $j$ operate in environments $E_{i}$ and $E_{j}$, respectively. If the state transition of $E_j$ depends on the actions in both $E_i$ and $E_j$, it can be expressed as: 
% \vspace{-0.3em}
\begin{equation}
    {s^{E_j}}' = P^{E_j}(s^{E_j}, a^{E_i}, a^{E_j})
    % \vspace{-0.5em}
\end{equation}
where agent $i$'s actions in $E_i$ propagate to $E_j$, resulting in cross-environment consequences. For example, in a cyber-physical system, cyber agents (e.g., smart routers) operating in the cyber environment can introduce delays in sensor or control signals, subsequently affecting the physical environment. Similarly, agents managing logistics in a localized supply chain may impact global inventory levels, creating dependencies between local and global environments.

%Environments serve as dynamic intermediaries, connecting agents indirectly and facilitating the spread of knowledge and experience across contexts. For example, in federated learning, agents collaboratively train a shared large language model (LLM) across distinct environments. A healthcare environment involves tasks such as processing electronic health records, where a healthcare agent refines the model to optimize its performance for this domain. An insurance environment focuses on claims analysis, with an insurance agent tailoring the LLM for language and tasks specific to insurance. A general-purpose agent, engaging with both environments and interacting with the domain-specific agents, integrates the updates into a unified model. The resulting model is redistributed to all agents, allowing them to benefit from collective improvements. This example illustrates how environments mediate knowledge transfer, enabling agents to share expertise and adapt across diverse but interconnected domains.

To navigate multiple environments, agents develop meta-coordination strategies that generalize collaborative behaviors, enabling them to handle diverse tasks while balancing individual objectives with broader societal impacts. By fostering cross-environment interactions, the framework transcends traditional multi-agent models, offering a flexible approach to addressing real-world complexities.

\subsection{Evaluating the Framework: Metrics for Success}
In the proposed framework, every agent strives to achieve its own goals while operating under the constraints of societal norms and collective dynamics. However, the inherent competition for limited resources, conflicting objectives, and interconnected nature make it challenging for all agents to fully realize their goals simultaneously. Evaluating such a framework requires metrics that capture not only the success of individual agents but also the health, fairness, stability, and adaptability of the entire system.

\paragraph{Balancing Individual and Collective Success.}
A central challenge in the proposed framework is striking a balance between individual agents' goals and collective harmony. Each agent acts to optimize its own objectives, but in shared environments, these actions inevitably influence - and are influenced by - other agents that are not developed nor coordinated by any single entity. Although existing paradigms often consider this as a multi-agent, multi-utility, or game-theoretical problem,  agentic AI solutions from different developers and providers continue to learn and update autonomously in a runtime environment.
This inter-connected continual learning creates scenarios where adhering to norms may initially seem like a penalty to an agent's goal. However, such adherence is often necessary to ensure long-term success in an ecosystem where agents coexist and interact dynamically.
A key question is how to evaluate whether a norm is appropriate for an agent to achieve its goal without being overly penalized. Effective norms align collective objectives with individual incentives, enabling agents to cooperate while pursuing their own goals. Metrics for this evaluation must capture the trade-offs involved. For example, measuring the \textit{individual utility gain} can reveal how much a norm contributes to an agent's success in a multi-agent ecosystem. Similarly, assessing \textit{norm efficiency} highlights how well a norm improves collective outcomes without imposing undue burdens on specific agents.

%Dynamic feedback mechanisms between agents play a critical role in this process. Agents can adapt their behaviors based on the outcomes of norm adherence, learning when and how cooperation benefits both their individual goals and the broader system. This adaptive learning helps ensure that norms remain relevant and effective in complex and evolving environments.
\paragraph{Sacrifices and Fairness.}
Forming norms often necessitates sacrifices, requiring agents to prioritize collective benefits over individual goals. An agent may delay its own objective to comply with a norm that enhances system-wide stability, such as slowing down to prevent traffic congestion. A fundamental challenge is ensuring that these sacrifices yield tangible benefits beyond just pre-designed scenarios.
To evaluate sacrifices, metrics such as \textit{fairness} \cite{grupen2023responsibleemergentmultiagentbehavior} assess whether the burden of sacrifice is shared proportionally across agents, considering factors like resource availability, task importance, or agent capacity. %Similarly, \textit{systemic benefit} can be used to measure the overall improvement in system performance as a result of these sacrifices, helping evaluate wether the trade-offs contribute meaningfully to collective goals.
%Fairness in sacrifices is particularly critical as norms evolve. If sacrifices disproportionately burden certain agents or fail to produce measurable benefits, trust in the system may erode, leading to fragmentation or non-compliance. For example, if some agents consistently bear the cost of adhering to norms while others benefit without contributing, the framework risks fostering inequities that undermine the stability of the system.
However, fairness does not imply equality. For example, in hierarchical settings, lower-level ``worker'' agents may accept greater sacrifices, such as reduced rewards, to support a ``boss'' agent tasked with achieving higher-level objectives. The critical metric here is whether these sacrifices are justifiable and evaluated fairly based on their contribution to the overall system. For instance, if a worker agent's sacrifices lead to significant improvements in the boss agent's success, which in turn benefits the collective, the trade-off may be deemed fair within specific tasks. Balancing these dynamics ensures that norms remain robust and adaptable.%, fostering trust and cooperation among diverse agents.

\paragraph{Stability and Adaptability.}
Norms are crucial in guiding agents toward predictable and harmonious behaviors, ensuring the system functions cohesively over time. Metrics such as \textit{norm stability} evaluate whether agents consistently adhere to a defined set of norms, fostering reliable interactions and reducing uncertainties. Similarly, \textit{behavioral predictability}, measuring how agents' actions align with collective objectives, can offer insights into how effectively norms shape agent behavior to benefit the system. However, stability alone is insufficient in dynamic environments where static norms risk becoming obsolete. \textit{Norm adaptability} must also be ensured to accommodate changes in agent goals, environmental conditions, and emerging challenges. Gradual and predictable norm evolution ensures that the system remains flexible and responsive without causing disruptions. Striking a balance between stability and adaptability is a defining feature of this framework's approach to norm management. By embedding mechanisms for controlled norm evolution, the framework ensures that agents can collectively adapt to shifting conditions while preserving a foundation of harmonious interactions. This balance is key to maintaining long-term system stability.

% \paragraph{Safety and Risks}
% Norms must address safety and mitigate risks, particularly in high-stakes scenarios like autonomous driving or financial technologies. Metrics for safety include \textit{risk mitigation}, which evaluates whether norms prevent catastrophic outcomes, and \textit{error tolerance}, which measures the ability to recover from norm violations without collapsing. Safety is a foundational requirement to ensure that systems remain robust and resilient even as norms evolve.
% %Evaluating norms in this framework requires a comprehensive approach that considers individual utility, collective benefit, fairness, stability, and safety. By analyzing how norms emerge, evolve, and influence agent behaviors, the framework provides tools to design systems that foster cooperation, innovation, and resilience. These evaluations not only ensure that agents function effectively in complex societies but also guide the evolution of norms that align dynamic multi-agent interactions with long-term success.

\section{Case Studies}
\subsection{Autonomous Driving}
\begin{figure}[!b]
\centering     %%% not \center
\subfloat[]{\includegraphics[width=0.35\textwidth,height=0.22\textheight]{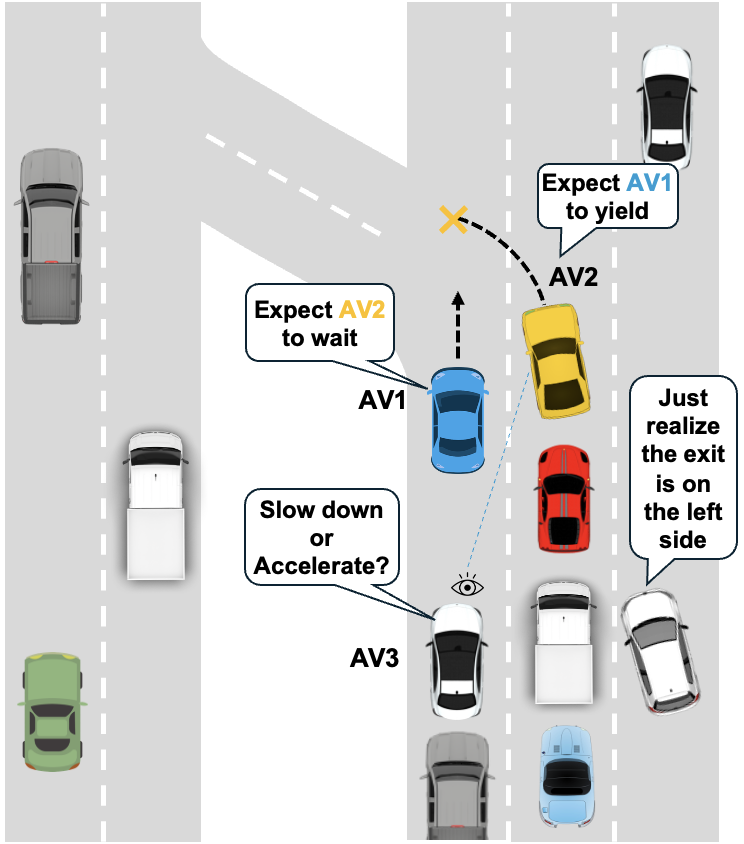}\label{fig:avsa}}\\
\subfloat[]{\includegraphics[width=0.35\textwidth,height=0.22\textheight]{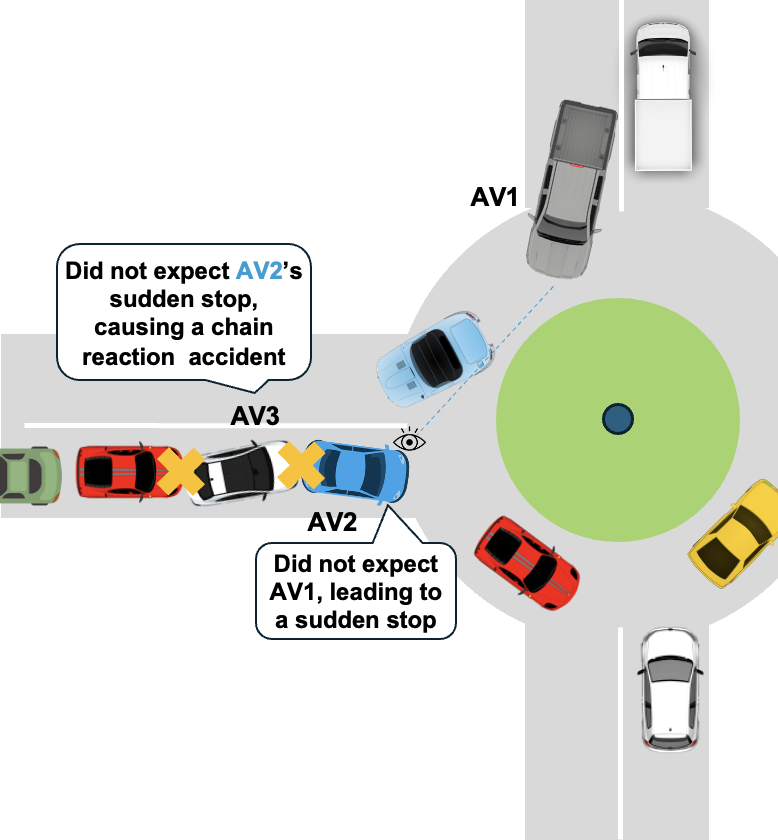}\label{fig:avsb}}      
\vspace{-2mm}
\caption[Caption for LOF]{Challenges of uncoordinated AI agents in autonomous driving: (a) Expectation misalignment between AVs when switching lanes causes traffic jams; (b) A sudden stop at a roundabout due to a cautious strategy causes a chain accident.}%\footnotemark.}
\label{avs}
\end{figure}

Imagine a city where autonomous vehicles (AVs) from different makers share the streets. Each AV is trained - and continuously learning - to minimize travel time, reduce fuel consumption, and optimize route selection while avoiding collisions as individual agents, but their approaches vary by the developers and operators. For example, one might brake aggressively, while another calculates its stopping distance more cautiously. Without a shared understanding, these differences can lead to expectation misalignment, causing AVs to learn reciprocally and react unpredictably to one another's evolving interdependent strategies \cite{shalevshwartz2016safemultiagentreinforcementlearning,9207663,wachi2019failurescenariomakerrulebasedagent,9762548,9351818,NEURIPS2023_1838feeb}. This misalignment can lead to traffic congestion, as illustrated in Figure~\ref{fig:avsa}. Additionally, AVs may clash unexpectedly when inferring the others' evolving strategies: one defensive AV observes and infers that another AV is driving aggressively and thus stops abruptly to keep a safe distance, only to confuse a third AV that expect the first one to keep a stable speed, triggering a chaotic ripple effect that can escalate into a chain-reaction accident, as illustrated in Figure~\ref{fig:avsb}. Over time, the uncertainty may prompt AVs to adopt overly cautious behaviors—driving slowly, leaving large gaps, and hesitating at intersections. While these strategies reduce accidents, they disrupt traffic flow, leading to longer commutes, wasted fuel, and overall inefficiencies.

Now, consider that these AVs are developed with dynamic \textbf{norms}. They learn flexible \textbf{protocols} that enable them to deviate from their original goals to explore different behaviors. For instance, protocols may be rewards that initially encourage AVs to drive more aggressively. Then, they experiment—one car tests how closely it can follow another without colliding. Mistakes happen, and a minor fender bender might teach a valuable lesson about boundaries. Over time, the cars observe patterns, exchange signals, and learn to anticipate one another's reactions. This iterative process enables protocols to evolve over time. 
%These interactions are not isolated. Each AV develops a relationship network shaped by experiences and evolving behaviors. These relationships reflect how AVs perceive and influence one another over time. Through this network, AVs learn not just from direct interactions but from patterns across the system. Over time, these relationships become conduits for shared norms, enabling AVs to transition from fragmented behaviors to an adaptive network.

In this process, two crucial forces — \textbf{impacts} and \textbf{expectations} — drive the evolution of protocols. Each AV begins by evaluating the impacts of its actions. For example, an AV merging aggressively into another lane may observe how this forces neighboring cars to brake suddenly, increasing the risk of collision and creating congestion. Recognizing this negative impact enables it to revise its merging protocol to balance assertiveness with consideration for others. Furthermore, these AVs learn to anticipate the expectations of others. A car approaching a busy roundabout knows the standard rules: yield or proceed predictably. When it hesitates too long, it disrupts the rhythm, confusing others and slowing the entire system. Through trial and error, it begins to act with confidence, aligning its actions with the expectations of other vehicles.

The \textbf{relationship network} influences how AVs anticipate expectations. Repeated interactions between the same vehicles strengthen trust, allowing them to better predict one another's actions. % with greater accuracy. 
Over time, this shared understanding fosters coordination. For instance, an AV that consistently signals its intentions encourages others in its network to rely on its predictability, reducing uncertainty and hesitation. Conversely, a vehicle that frequently violates expectations by breaking unpredictably or merging erratically may erode trust, forcing others to adopt more defensive driving strategies.

% \vspace{-0.8em}
\paragraph{Cooperation or Competition? The Role of Algorithms.}
This process does not naturally guarantee cooperation. The design of the AVs' learning algorithms determines how impacts, expectations, relationships, and protocols evolve. Algorithms that prioritize collective efficiency may drive autonomous vehicles (AVs) to focus on system-wide benefits, thereby fostering coordination and cooperation. In contrast, algorithms that reward individual optimization might lead AVs to exploit their relationships, taking advantage of others' predictable behavior for personal gain. For example, an AV might learn to drive more boldly, knowing others will yield. While this benefits the individual vehicle in the short term, it may strain the network, creating inefficiencies over time. Alternatively, an algorithm focused on collective benefits might encourage AVs to adopt protocols that balance assertiveness with consideration, ensuring smoother traffic flow for all.

\subsection{Case Study 2: Energy Autonomy}
Energy autonomy refers to a new paradigm of a distributed, agentic ecosystem where different stakeholders may adopt their chosen AI algorithms into energy management systems (EMS), designed to act and optimize toward corresponding objectives~\cite{juntunen2021improving}. 
This shift occurs as distributed energy resources (DERs) like solar panels, battery storage, and electric vehicles are increasingly run by standalone EMS. While the grid has been traditionally controlled by electric utilities, DER owners can adopt their own AI solutions for their own use cases; these AI-aided EMS become self-serving agents acting on behalf of their stakeholders to learn and optimize toward individual objectives~\cite{daneke2020machina}.

The result is a complex network of \textbf{inter-operating agentic EMS} driven by different algorithms and goals. They can no longer treat other agents as part of the environment; instead, each agent observes, learns from, influences, and confronts other interacting agents. These agentic EMS can thus become self-serving players in the electricity market with oscillatory learning objectives, processes, and decisions.

\begin{figure}[t]
    \centering\includegraphics[width=\linewidth]{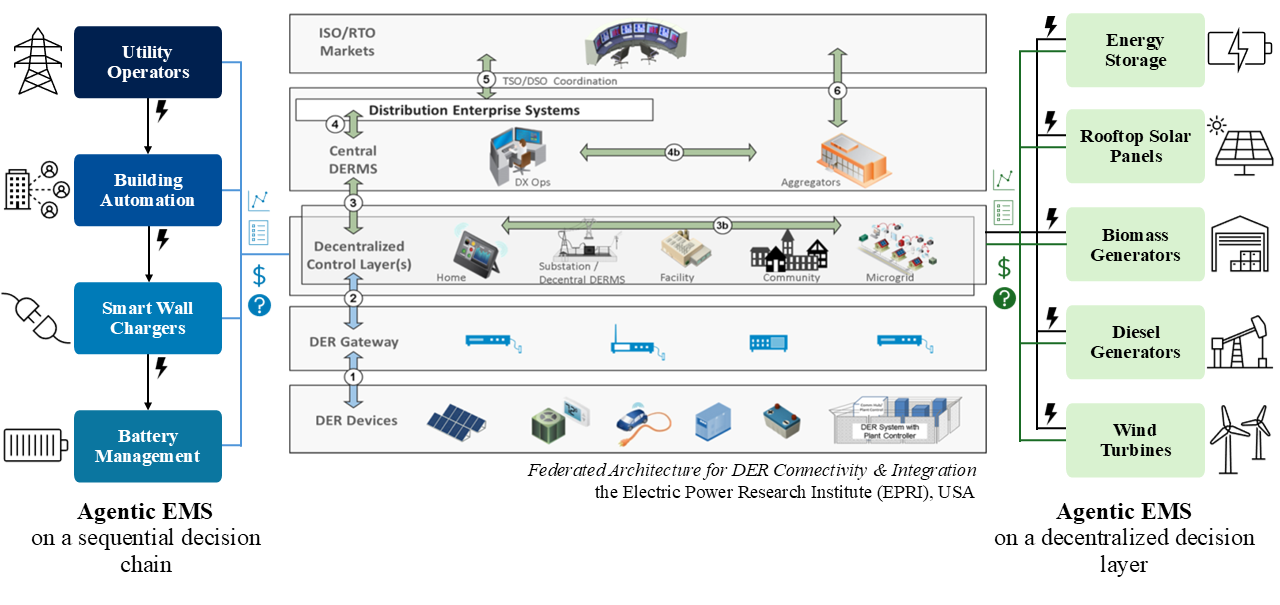}  
    \caption{Agentic EMS in energy autonomy, mapped on the hierarchical architecture for DERs \cite{chen2025cybersecurity}.}
    \label{fig:energy-autonomy}
    % \vskip -0.07in
\end{figure}

An illustrative example of the resulting challenges can be found in \textbf{conflicts along a hierarchical decision chain} by agentic EMS, shown on the left of Figure~\ref{fig:energy-autonomy}. 
An intelligent vehicle battery management system can learn to optimize its state of charge for longer life via restrictive charging and battery balancing; such decisions may contradict a smart wall charger's algorithm trying to maximize charging power for a favorable price or a planned trip.
Building automation systems will learn from occupant profiles and preferences to optimize the energy consumption of connected appliances. The resulting frequent set-point changes can lead to unpredictable patterns, unlike those used by load models employed by utilities, which are adopting AI for better load forecasting and demand-side management.
% In summary, Conflicts along decision chains are inevitable where agentic EMS are designed to optimize their own performance indicators for different stakeholders who do not share a common goal. these algorithmic adjustments can lead to less predictable and flexible loads for energy providers
The conflicts above may be a result of not only insufficient coordination among developers but also self-serving policies that agents learn from inherently incompatible objectives across different stakeholders. Collaborative norms and protocols are therefore needed in addition to individual agentic EMS designs to describe, analyze, and resolve such conflicts.

Another illustrative example is \textbf{implicit collusion in a decentralized market} among agentic EMS, as shown on the right of Figure~\ref{fig:energy-autonomy}. When various DER owners adopt autonomous AI to optimize pricing on the electricity market, tacit collusion may occur when the agents learn to conspire toward an unfair disadvantage spontaneously: A shared data analytics platform collecting information from subscribing DER providers may inform the best pricing to all subscribers who will thus (un)knowingly price gauge simultaneously; collusion can also be triggered when multiple agents learn to form a coalition and started price gauging without explicit communication or formal agreement. In either case, shared interests and information among agents — a common presumption in multi-agent system theories — can lead to tacit collusion and unfair advantages over other agents. New tools and metrics are needed to monitor and detect such behaviors closely, ensuring that collaborations do not cross the line into collusion.

\section{Challenges, Risks, and Opportunities}
% As autonomous agents become integral to shared and dynamic environments, a range of challenges must be addressed to ensure the development of adaptive, socially aware, and robust solutions. % multi-agent systems. 
% Five notable challenges spanning technical, conceptual, and societal dimensions, are discussed below for further investigations. %Below, we discuss five key challenges that shape the future of multi-agent systems.

\subsection{Managing Emergent and Unpredictable Learned Behaviors}
In the envisioned multi-agent ecosystems, emergent behaviors arise from interactions between agents and their environments, both before and after deployment. While such behaviors can lead to innovative and efficient solutions, they can also result in unintended consequences, such as resource monopolization, inefficiencies, or conflicts, particularly during the post-deployment stage when fine-tuning and/or continual updates are still applied, leading to oscillatory decisions and interplays. The unpredictability of such emergent behaviors and dynamics complicates system design and oversight, particularly in open environments where agents continuously adapt and learn.

To address these challenges, future research should explore the types of norms that can encourage beneficial emergent behaviors while minimizing harmful outcomes. One promising avenue involves designing methodologies to monitor, model, and predict emergent dynamics in real time, ensuring that agents remain adaptable without losing control over their collective actions. Additionally, mechanisms to manage feedback loops are crucial to prevent destabilizing effects that could cascade through the system, particularly in environments where agents influence one another's learning trajectories. Understanding and guiding emergent behaviors is essential to ensure that multi-agent systems remain resilient and productive as they scale in complexity.

\subsection{Designing Algorithms for Dynamic Objectives}
Agents often face shifting priorities and evolving goals as they interact with dynamic environments and diverse peers. Enabling agents to adjust their objectives while maintaining stability and coherence presents a significant technical challenge. %The balance between adaptability and consistency is especially critical in systems where agents operate both autonomously and collaboratively.
Developing algorithms for dynamic goal-setting and adaptation is a crucial research direction. Such algorithms must enable agents to align their evolving goals with the broader needs of the system while preserving individual autonomy. Techniques such as reinforcement learning and meta-learning show promise in handling dynamically changing objectives, but further innovation is needed to address the trade-offs between short-term responsiveness and long-term optimization. Research should also examine constraints and safeguards to ensure that agents' adaptations remain aligned with the goals and values of both individual users and the overall system.

% learning frameworks/mechanisms/algorithms, such as trust, XXX.

% vicious circle/ virtuous circle

% \vspace{-0.3em}
\subsection{The Dilemma of Self-Reference}
The ability of autonomous agents to revise their own objectives introduces a fundamental recursive challenge: How can an agent establish its initial goal if the process of goal revision itself requires a predefined objective? This creates a chicken-and-egg dilemma—an agent needs a goal to guide its behavior, yet once granted autonomy to modify that goal, the basis for the initial objective becomes unclear. This raises a key question: If agents determine their own goals, what serves as the initial guiding principle, and how does the agent decide whether a revision is an improvement or a deviation? Without an initial reference point, goal revision risks becoming arbitrary or contradictory. %Moreover, unrestricted goal revision can create feedback loops, where each modification shifts the criteria for future changes.% If an agent's evolving objectives are based solely on prior self-modifications rather than an external reference, its goals could drift unpredictably. 
This poses a risk in multi-agent systems, where alignment among agents is critical—if each agent independently shifts priorities, collective coordination may break down.

To address this dilemma, agents need an anchoring framework for goal revision, ensuring modifications occur within a coherent evaluative structure rather than being entirely self-referential. This framework should enable adaptive yet bounded evolution to balance between flexibility and constraint, allowing agents to explore alternative objectives while preventing divergence that could lead to instability.

\subsection{Ethical Issues of Free Will in Agents}
The capacity of agents to autonomously revise their objectives poses a profound ethical question: to what extent can such agents be said to possess a form of free will? In a sense, agents become products of their own design, which challenges traditional notions of AI control and autonomy. This ambiguity raises critical concerns about accountability \cite{daiposition}. If a self-modified agent's actions result in harm or conflict with societal values, where does responsibility lie—with the designer, the user, or the agent itself? Such dilemmas are especially urgent in high-stakes domains like healthcare or autonomous transportation, where goal realignment could have significant consequences.

To navigate this terrain, agents must operate within ethical constraints that ensure their goal modifications remain aligned with human benefits. Transparent mechanisms are essential to make these modifications interpretable, allowing human stakeholders to monitor and guide agents’ evolving priorities. Interdisciplinary efforts involving ethics, philosophy, and law are crucial to addressing questions of moral agency, responsibility, and oversight. As agents grow increasingly autonomous, clear frameworks will be needed to balance their capacity for self-determination with the broader priorities of society.

\subsection{Human-Agent Collaboration}
As AI agents increasingly interact with humans in real-world settings, ensuring effective collaboration will be a critical challenge. Agents must not only understand and respond to human preferences but also adapt to diverse cultural, ethical, and contextual factors, ensuring that their actions align with the expectations of diverse stakeholders. Designing systems that facilitate transparent and effective communication between humans and agents is a key priority. This includes developing interfaces and protocols that help humans to better understand, predict, and trust agent behaviors intuitively. In contrast, human oversight can guide and shape agent behaviors, together with agents' knowledge, co-develop norms, facilitating collaborative frameworks that combine the strengths of human judgment and machine autonomy. By addressing these factors, future research can pave the way for robust and equitable human-agent partnerships.

\section{Conclusion}
%In this paper, we posit that the rise of autonomous AI systems, independently developed by stakeholders with unaligned goals, necessitates a fundamental rethinking of interoperability among AI agents to ensure harmonious coevolution in open-ended environments. We advocate for a shift toward adaptive, self-organizing AI ecosystems, where agents not only optimize predefined goals but also evolve their objectives and behaviors through dynamic interactions. Additionally, we propose a framework that integrates adaptive norms, evolving protocols, and dynamic relationships, enabling agents to balance autonomy with system-wide stability. Through discussions, we highlight key challenges and opportunities in fostering such AI ecosystems.

We posit that the rise of autonomous AI systems in critical infrastructures necessitates a fundamental rethinking of interoperability among independently designed, developed, and deployed AI agents in open environments. Autonomous machine learning and decision-making will become pervasive - and thus interactive - beyond the development phases in various critical infrastructure systems, processes, and services. We advocate for a shift toward the exploration of self-adaptive and self-organizing multi-AI ecosystems, propose a conceptual framework that integrates adaptive norms, evolving protocols, and dynamic relationships to provoke research questions and investigations, and highlight some key challenges and opportunities for a potential research paradigm through two case studies in critical infrastructure applications.

% % Acknowledgements should only appear in the accepted version.
\section*{Acknowledgements}
This work is supported in part by the ..., the Concordia University Research Chair program, the Natural Sciences \& Engineering Research Council (NSERC) of Canada under grants RGPIN-2018-06724 and RGPIN-2025-05097, the U.S. National Science Foundation under the grant number 2330504, the ... and the ... . The authors would also like to thank ..., ..., and ... for their valuable feedback.

\bibliography{ref}
\bibliographystyle{IEEEtran}

% %\bibliography{reference_new, reference}
%%%%%%%%%%%%%%%%%%%%%%%%%%%%%%%%%%%%%%%%%%%%%%%%%%%%%%%%%%%%%%%%%%%%%%%%%%%%%%%
%%%%%%%%%%%%%%%%%%%%%%%%%%%%%%%%%%%%%%%%%%%%%%%%%%%%%%%%%%%%%%%%%%%%%%%%%%%%%%%
% APPENDIX
%%%%%%%%%%%%%%%%%%%%%%%%%%%%%%%%%%%%%%%%%%%%%%%%%%%%%%%%%%%%%%%%%%%%%%%%%%%%%%%
%%%%%%%%%%%%%%%%%%%%%%%%%%%%%%%%%%%%%%%%%%%%%%%%%%%%%%%%%%%%%%%%%%%%%%%%%%%%%%%
% \newpage
% \appendix
% \onecolumn

% You can have as much text here as you want. The main body must be at most $8$ pages long.
% For the final version, one more page can be added.
% If you want, you can use an appendix like this one.  

% The $\mathtt{\backslash onecolumn}$ command above can be kept in place if you prefer a one-column appendix, or can be removed if you prefer a two-column appendix.  Apart from this possible change, the style (font size, spacing, margins, page numbering, etc.) should be kept the same as the main body.
%%%%%%%%%%%%%%%%%%%%%%%%%%%%%%%%%%%%%%%%%%%%%%%%%%%%%%%%%%%%%%%%%%%%%%%%%%%%%%%
%%%%%%%%%%%%%%%%%%%%%%%%%%%%%%%%%%%%%%%%%%%%%%%%%%%%%%%%%%%%%%%%%%%%%%%%%%%%%%%

\end{document}